\documentstyle[12pt]{article}
\begin{document}
\input epsf

\def\figresall
{
\begin{figure}[tbh]
\epsfysize=4.0in
\centerline{\epsfbox{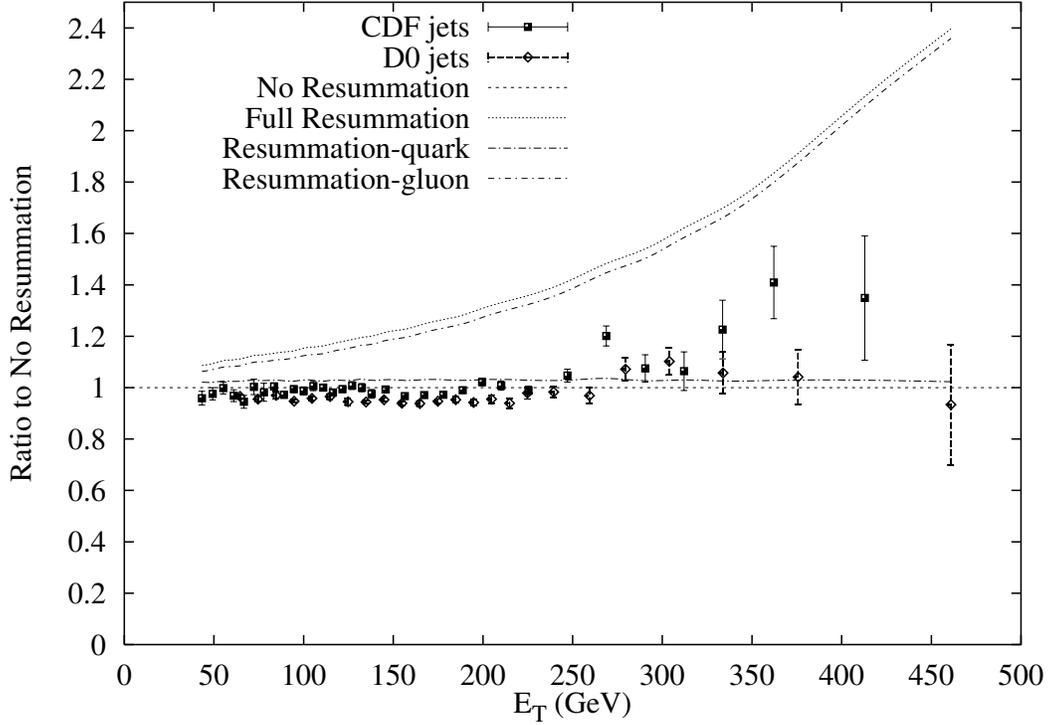}}
\caption{Enhancements from the full threshold resummation,
from the resummation for quark only, and from the resummation for
gluon only. The NLO predictions and experimental data 
(with statistical errors only) are also shown.}
\label{fig:resall}
\end{figure}
}

\def\figresglu
{
\begin{figure}[tbh]
\epsfysize=4.0in
\centerline{\epsfbox{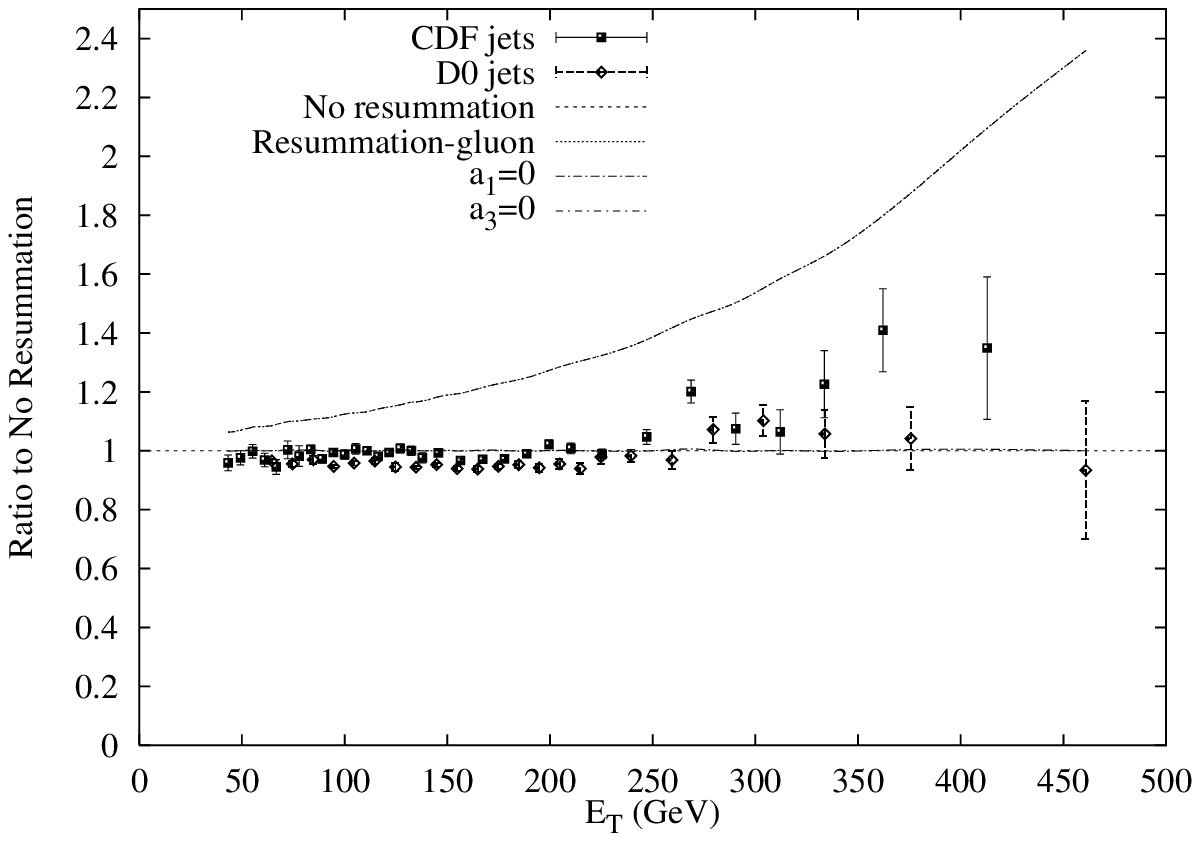}}
\caption{ Enhancements from the threshold resummation for gluon,
from the resummation with $a_1$ set to zero, and from the resummation
with $a_3$ set to zero (which appears to coincide with dotted line for gluon). 
The NLO predictions and experimental data
(with statistical errors only) are also shown.}
\label{fig:resglu}
\end{figure}
}

\hfill{NCKU-HEP-99-07}\par
%\hfill{hep-ph/9906XXX}
\vskip 0.3cm
\begin{center}
{\large {\bf Effects of threshold resummation}}
\vskip 1.0cm
Hung-Liang Lai\footnote{lai@phys.nthu.edu.tw}
\vskip 0.3cm
Department of Physics, National Tsing-Hua University, \par
Hsinchu, Taiwan 300
\vskip 0.3cm
Hsiang-nan Li\footnote{hnli@mail.ncku.edu.tw}
\vskip 0.3cm
Department of Physics, National Cheng-Kung University, \par
Tainan, Taiwan 701, Republic of China
\vskip 0.3cm
Theory Group, KEK, Tsukuba, Ibaraki 305, Japan
\end{center}
\vskip 1.0cm

%PACS numbers: 12.38.Bx, 12.38.Cy
\vskip 1.0cm
%\baselineskip=2\baselineskip

\centerline{\bf Abstract}
\vskip 0.3cm

We investigate effects of threshold resummation of logarithmic
corrections $\ln N$ in Mellin space quantitatively. Threshold resummation 
leads to enhancement of next-to-leading-order QCD predictions for jet 
production at large jet transverse energy, which is in the trend indicated
by experimental data. We show that this enhancement is completely 
determined by the behavior of threshold resummation at small $N$, the 
region where hierachy among different powers of $\ln N$ is lost and current 
next-to-leading-logarithm resummation is not reliable. Our analysis 
indicates that more accurate threshold resummation formalism should be 
developed in order to obtain convincing predictions.

%\newpage

\section{INTRODUCTION}

The formalism of threshold resummation of double logarithmic corrections
to QCD processes, which occur in extreme kinematic conditions, has been
developed for some time \cite{S0,CT,KM}. At the kinematic end points, a 
special type of corrections $\ln(1-x)/(1-x)_+$ is produced with $x$ being 
a parton momentum fraction, which appears as $\ln^2(1/N)$ under the Mellin 
transformation. There have been abundant formal derivations of 
threshold resummation for various processes, such as deep inelastic
scattering, Drell-Yan, direct photon, and heavy-quark productions
\cite{S}. Quantitative studies of threshold resummation effects in 
heavy quark production \cite{KV} and in direct 
photon production \cite{CM} have been performed recently. Such 
numerical studies are essential in order to justify that threshold resummation 
indeed collects important dynamics of processes in extreme kinematic 
conditions.

In this letter we shall analyze effects of threshold resummation from another 
point of view. For our purpose, it suffices to consider resummation 
of the logarithmic corrections that can be factorized into parton
distribution functions (PDFs). Taking jet production at Tevatron as an example
\cite{data}, we observe that threshold resummation enhances 
next-to-leading-order (NLO) QCD predictions for jet production at large jet 
transverse energy $E_T$. This tendency is qualitatively consistent with 
experimental data and with the conclusion in \cite{KV,CM}. However, the 
enhanced predictions overshoot data by a factor of 2. A simple investigation 
reveals that the resummation associated with the quark distribution 
functions lead to a negligible effect and the overestimation is attributed to 
the resummation associated with the gluon distribution function. The reason 
is that the color factor $N_c=3$ in the latter case is larger than 
$C_F=4/3$ in the former case. 

We further find that the enhancement is completely determined by the 
behavior of the resummation associated with the gluon distribution function 
at low $N$. Unfortunately, this is a region where current 
next-to-leading-logarithm (NLL) resummation is  
not reliable, since hierachy among different powers of $\ln N$ is lost 
and all nonleading logarithms need to be summed. In other words, the 
large-$N$ behavior of threshold resummation is reliable, but almost
irrelevant to the end-point enhancement, whereas the small-$N$ 
behavior accounts for the end-point enhancement, but is not reliable in
NLL threshold resummation. Moreover, the 
importance of low-$N$ contributions can not be diminished no matter how 
extreme kinematic conditions are. Our analysis indicates that more accurate 
threshold resummation formalism need to be developed in order to obtain 
convincing predictions.

\section{THRESHOLD RESUMMATION}

Threshold resummation of the logarithmic corrections that can be 
factorized into PDFs is written as \cite{S}
\begin{eqnarray}
{\tilde f}(N)=\exp\left[\int_{0}^{1}dz\frac{1-z^{N-1}}{1-z}
\int_{(1-z)^2}^{1}\frac{d\lambda}{\lambda}
\gamma_{K}(\alpha_s(\sqrt{\lambda} p^+))\right]\;,
\label{fbt2}
\end{eqnarray}
where $p^+$ is the longitudinal component of hadron momentum $p$, and
the anomalous dimension $\gamma_K$ is given, up to two loops, 
by \cite{CS}
\begin{equation}
\gamma_K=\frac{\alpha_s}{\pi}C_F+\left(\frac{\alpha_s}{\pi}
\right)^2C_F\left[C_A\left(\frac{67}{36}
-\frac{\pi^{2}}{12}\right)-\frac{5}{18}n_{f}\right]\;,
\label{lk}
\end{equation}
with $C_A=3$ being a color factor and $n_{f}=4$ the number 
of quark flavors. In this work we shall adopt its modified version,
\begin{eqnarray}
{\tilde f}(N)=\exp\left[\int_{0}^{1-1/N}\frac{dz}{1-z}
\int_{(1-z)^2}^{1}\frac{d\lambda}{\lambda}
\gamma_{K}(\alpha_s(\sqrt{\lambda} p^+))\right]\;,
\label{fbt}
\end{eqnarray}
which is equivalent to Eq.~(\ref{fbt2}) up to $O(1/N)$ corrections. We
refer readers to \cite{L} for its detailed derivation. Equation 
(\ref{fbt}) is simpler in analytical manipulation, since the integral
in the exponent can be worked out explicitly. Threshold resummation 
associated with the gluon distribution function is obtained by substituting
$N_c=3$ for $C_F$ in Eq.~(\ref{lk}).

The first term of $\gamma_K$ leads to the leading (double)-logarithm
summation, and the second term leads to the NLL summation. Note that 
Eq.~(\ref{fbt}) is not complete at the NLL level. To obtain a full NLL 
summation, contributions from soft gluon exchanges among initial- and 
final-state partons should be taken into account, which are 
process-dependent. While Eq.~(\ref{fbt}) is process-independent, since it
sums factorizable corrections. In this work we shall comment on 
applications of threshold resummation to various QCD processes, such as 
deep inelastic scattering, direct photon production, and jet production.
Furthermore, our goal is to demonstrate that the end-point enhancement is 
determined by the small-$N$ behavior of threshold resummation. Hence, 
Eq.~(\ref{fbt}) serves the purpose.

We propose an expansion of the threshold resummation ${\tilde f}(N)$ 
in Eq.~(\ref{fbt}) in terms of polynomials in $N$:
\begin{equation}
{\tilde f}(N)=\sum_{i=0}^n a_iC(N-1,i)\;,\;\;\;\;
C(N,i)\equiv \frac{N!}{i!(N-i)!}\;.
\label{fm}
\end{equation}
The above series corresponds to a simple form in momentum fraction space,
\begin{equation}
f(x)=\sum_{i=0}^n \frac{a_i}{i !}\delta^{(i)}(1-x)\;,
\label{mx}
\end{equation}
which can be easily verified by performing the Mellin transformation
\begin{equation}
{\tilde f}(N)=\int_0^1 dx x^{N-1} f(x)\;.
\end{equation}
The first coefficient $a_0={\tilde f}(1)=1$ gives the initial condition of 
threshold resummation. The other coefficients $a_i$ are determined by 
best fit to Eq.~(\ref{fbt}). On the other hand, Eq.~(\ref{fbt}) is not 
appropriate for large $N$, the region in which the integration variable 
$\lambda$  may be as small as $1/N^2$, the running coupling constant
$\alpha_s(p^+/N)$ diverges, and perturbation theory is not applicable. 
In this region Eq.~(\ref{fbt}) should be replaced by a nonperturbative 
function, which is of course model-dependent. For example, a minimal
prescription which takes into account only the Landau singularity was
introduced in \cite{CMN}. It is straightforward to extrapolate 
Eq.~(\ref{fm}) to the $N\to \infty$ limit, and this extrapolation can be
regarded as a nonperturbative model.

It turns out that an expansion up to $n=3$ in Eq.~(\ref{fm}) describes the 
growth of Eq.~(\ref{fbt}) with $N$ very precisely. 
The parameters $a_i$ for the quark and gluon distribution functions 
from best fit to Eq.~(\ref{fbt}) for $\Lambda_{\rm QCD}=0.2$ GeV 
in $\alpha_s$ and for center-of-mass energy $\sqrt{s}=\sqrt{2}p^+=1800$ GeV
at Tevatron are listed below:
\begin{center}
\begin{tabular}{cccc}
\hline
    parton & $a_1$ & $a_2$ & $a_3$  \\
\hline
quark &$1.6198\times 10^{-2}$&$-8.5872\times 10^{-6}$&$2.2515\times 10^{-8}$ \\
gluon &$1.1248\times 10^{-1}$&$7.2253\times 10^{-5}$&$2.5192\times 10^{-6}$ \\
\hline
\end{tabular}
\end{center}
For smaller $\sqrt{s}$, such as $\sqrt{s}=38.7$  GeV for direct photon 
production in E706 \cite{E706}, we obtain the parameters
\begin{center}
\begin{tabular}{cccc}
\hline
    parton & $a_1$ & $a_2$ & $a_3$  \\
\hline
quark &$5.6075\times 10^{-2}$&$6.6109\times 10^{-4}$&$-2.1474\times 10^{-6}$ \\
gluon &$1.3616\times 10^{-1}$&$9.3512\times 10^{-3}$&$5.7660\times 10^{-4}$ \\
\hline
\end{tabular}
\end{center}
The parameters $a_i$ increase as $\sqrt{s}$ decreases, implying stronger 
resummation effects, because the running $\alpha_s$ is larger at lower 
energies. For even lower $\sqrt{s}$, such as those for deep inelastic 
scattering, $a_i$ are even larger.

The modified PDF $\bar\phi$ is written as the 
convolution of threshold resummation with the original distribution function 
$\phi$ \cite{S}:
\begin{eqnarray}
{\bar\phi}(x)&=&\int_x^1\frac{d\xi}{\xi}f(\xi)\phi(x/\xi)\;,
\nonumber\\
&=&(1-a_1+a_2-a_3)\phi(x)-(a_1-2a_2+3a_3)x\frac{d}{dx}\phi(x)
\nonumber\\
& &+\frac{1}{2}(a_2-3a_3)x^2\frac{d^2}{dx^2}\phi(x)
-\frac{1}{6}a_3x^3\frac{d^3}{dx^3}\phi(x)\;.
\label{mpdf}
\end{eqnarray}
Briefly speaking, threshold resummation effectively modifies a PDF, and the 
modification is energy- and process-dependent.
Hence, before performing a global determination of PDFs,
one should clarify threshold resummation effects.

The motivation to expand the threshold resummation into a series of 
$C(N-1,i)$ up to $i=3$ is as follows. The $C(N-1,1)$ term and the 
$C(N-1,3)$ term determine the behavior of the resummation at small 
$N$ and at large $N$, respectively. If the series terminates at $i< 3$, 
the role of each $C(N-1,i)$ in determining the behavior of the
resummation in different regions of $N$ is not significant. If the series 
contains terms with $i >3$, numerical handling of higher derivatives of 
modified PDFs will be difficult. Since the 
resummation associated with the gluon distribution function dominates, we 
take it as an example to demonstrate the above idea. For $N\sim 10$, the 
$i=1$ term $a_1C(N-1,1)$ in the case with $\sqrt{s}=1800$ GeV is of order 
unity, while the $i=3$ term $a_3C(N-1,3)$ is only of order $10^{-3}$. As 
$N$ increases up to $10^3$, the edge for perturbation theory to be 
applicable, the $i=1$ term, being of order $10^2$, becomes smaller than the 
$i=3$ term, which is of order $10^3$. Hence, a variation of $a_1$ implies a 
variation of the small-$N$ behavior of threshold resummation, and a 
variation of $a_3$ implies a variation of the large-$N$ behavior.

For $N\sim 10^3$, a double logarithm $\ln^2 N$ is larger than a single 
logarithm $\ln N$ by a factor of 7, indicating that hierachy among different 
powers of $\ln N$ exist and current NLL resummation is reliable. For $N$ as 
small as 10, $\ln^2 N$ and $\ln N$ are in fact of the same order, and 
NLL resummation is not reliable. Therefore, by varying the coefficients 
$a_1$ and $a_3$, we can investigate how sensitive the end-point enhancement 
of NLO predictions is to the small-$N$ portion of the NLL resummation, which 
is not reliable, and to the large-$N$ portion, which is reliable. 
It will be satisfactory, if the enhancement is insensitive to the 
small-$N$ behavior of the NLL resummation. However, we shall demonstrate
that this is not the case. The reason is obvious from Eq.~(\ref{mpdf}):
the coefficient of each term on the right-hand side of Eq.~(\ref{mpdf})
is dominated by $a_i$ with smaller $i$. The above reasoning applies to cases 
like E706 with $\sqrt{s}=38.7$ GeV or lower, for which the corresponding
parameters $a_i$ have the same relation $a_1\gg a_2\gg a_3$. In these cases
the end-point enhancement is determined by the behavior of threshold
resummation at smaller $N$, and the controversy stated above is more
serious.

\section{NUMERICAL ANALYSIS}

Consider jet production with transverse energy $E_T$ in $p{\bar p}$
collision,
\begin{equation}
p(p_1)+{\bar p}(p_2) \to J(E_T) + X\;.
\label{dy}
\end{equation}
The hadron momenta are assigned as $p_1=(p_1^+,0,{\bf 0}_T)$ and 
$p_2=(0,p_2^-,{\bf 0}_T)$ with $p_1^+= p_2^-=\sqrt{s/2}$. Partons carry the 
momenta $\xi_i p_i$ with $\xi_i$, $i=1$, 2, being the momentum fractions. 
In fact, the transverse momenta ${\bf k}_{iT}$ of partons should be taken 
into account, when transverse degrees of freedom of final states are 
measured \cite{LL}. The corresponding $k_T$ resummation, if included, is 
expected to further enhance the cross section at high $E_T$. 
In the present work $k_T$ resummation will not be considered, since we
concentrate on effects of threshold resummation.

The factorization of jet production is basically similar to that of
direct photon production in \cite{LL}. The self-energy correction to a 
parton and the loop correction with a real gluon connecting the two partons 
from the same hadron, contain both collinear divergences from the loop 
momentum $l$ parallel to $p_i$ and soft divergences from small $l$. Since 
soft divergences cancel between the above corrections \cite{L1}, the 
remaining collinear divergences are absorbed into a PDF
associated with the hadron $i$. They are the corrections which 
produce the logarithms $\ln (1/N)$ that have been summed into 
Eq.~(\ref{fbt}) in axial gauge. In the considered kinematic region for jet 
production, the other radiative corrections, because of the soft 
cancellation, are absorbed into a hard scattering amplitude $H$, which 
corresponds to a parton-level differential cross section. 

The factorization formula for jet production with the threshold resummation
for PDFs included are written as
\begin{eqnarray}
\frac{d\sigma(E_T)}{dE_T}=
\int d\xi_1 d\xi_2 
{\bar\phi}(\xi_1,E_T/2){\bar\phi}(\xi_2,E_T/2)
H(\xi_1,\xi_2,s,E_T/2)\;.
\label{fdk}
\end{eqnarray}
We have set the renormalization (factorization) scale $\mu$ of
$\bar\phi$ and $H$ to the characteristic scale $E_T/2$. The original 
PDFs $\phi$ evolve to $E_T/2$ according to the
Dokshitzer-Gribov-Lipatov-Altarelli-Parisi equation \cite{AP}, which
sums another type of single logarithms $\ln E_T$. In the following numerical 
analysis we employ the NLO QCD calculations for jet production derived 
in \cite{EKS} and the CTEQ4M set \cite{cteq4} for the orginal  
PDFs. We have also checked the CTEQ3M set
\cite{cteq3} and found that our conclusion does not depend on the choice of 
PDF sets. 

\figresall
In Fig.~1 we show the modification from the threshold resummation on NLO 
QCD predictions for jet production at Tevatron. It has been observed
that there seems to be an excess of the CDF data at high $E_T$
compared to the NLO predictions with usual PDFs \cite{cteq4,cteq3,mrs,cteq5},
whereas the D0 data are in good agreement with the predictions.
Note that the CDF and D0 data do not conflict each other if
considering the large systematic uncertainties ($8\sim 30\%$)
in addition to the statistical ones. 
Hence, it is expected that the threshold resummation causes a small amount 
of enhancement of the NLO predictions at high $E_T$. However, it is found 
that the enhancement is a factor of 2. We explore the source that is 
responsible for this huge overestimation. If turning off the threshold 
resummation associated with the gluon distribution function, the 
end-point enhancement falls dramatically. If turning off the threshold 
resummation associated with the quark distribution function, the 
enhancement almost remains invariant. That is, the resummation associated
with the quark distribution function contributes only few percent of the 
full enhancement, and the overestimation is attributed to 
the resummation associated with the gluon distribution function as shown in 
Fig.~1.

\figresglu
We then investigate which portion in $N$ of the resummation associated with
the gluon distribution function accounts for the end-point enhancement. As 
stated before, the parameters $a_1$ and $a_3$ control the small-$N$ and 
large-$N$ behaviors of threshold resummation, respectively. 
If setting $a_1$ to zero, which 
changes the small-$N$ behavior of the resummation but leaves almost
invariant the large-$N$ behavior, the resultant predictions coincide with 
those from the NLO calculations without including resummation
effects as shown in Fig.~2. If setting $a_3$ to zero, which 
changes the large-$N$ behavior but leaves invariant the small-$N$ behavior,
the enhancement remains the same. This investigation indicates that the 
behavior of the resummation at small-$N$ determines the end-point enhancement. 
As argued before, this is the region where hierachy among different powers 
of $\ln N$ is lost and more accurate formulas including summation of all
nonleading logarithms are required. Before extracting reliable predictions 
from threshold resummation, this point must be taken into account. 
At last, we observe that the end-point enhancement can be adjusted by 
tuning $a_1$, {\it i.e.}, by varying the small-$N$ behavior of threshold 
resummation. Choosing $a_1=0.02$ in the resummation associated with the 
gluon distribution function, the resultant predictions are similar to those 
from applying the CTEQ4HJ PDFs \cite{cteq4,cteqj}, and well describe
the CDF and D0 data simultaneously. 

We emphasize that the above controversy always exists no matter how large
$E_T$ is reached. With higher $E_T$, behaviors of  
PDFs at larger momentum fraction are probed. In this region
we have $x\to 1$ in the inverse Mellin transformation,
\begin{equation}
f(x)\equiv\frac{1}{2\pi i}\int_{c-i\infty}^{c+\infty} 
dN x^{-N}{\tilde f}(N)\;,
\label{inm}
\end{equation}
with $c$ being an arbitrary constant,
and contributions from ${\tilde f}(N)$ in the whole range of $N$ are 
equally important. In cases with intermediate $E_T$, for which $x$ may not 
approach unity, contributions from the small-$N$ region dominate. That 
is, no matter how large $E_T$ is reached, the small-$N$ 
region always contributes. We have also analyzed the data of direct
photon production in E706 \cite{E706} and of deep inelastic scattering in
BCDMS \cite{bcdms} and NMC \cite{nmc}
with lower $\sqrt{s}$ using the above method, and arrived at the same 
conclusion: in the region where perturbation theory is applicable, 
small-$N$ contributions always determine the end-point enhancement.

\section{CONCLUSION}

In this letter we have shown that the behavior of the threshold resummation 
associated with the gluon distribution function at small $N$ determines 
the end-point enhancement of NLO predictions for jet production completely. 
A variation of the large-$N$ behavior of the NLL resummation, which is
reliable, does not affect the enhancement. However, in the small-$N$ region 
hierachy among different powers of $\ln N$ disappears and current NLL 
resummation is not reliable. This controversy also exists in processes
with higher $E_T$ or lower center-of-mass energies. Hence, to obtain 
convincing predictions, more accurate formalism for threshold resummation 
which sums all nonleading logarithms need to be developed. We emphasize that 
it is not the goal of this work to explain experimental data. Even if data
can be explained by complete NLL threshold resummation, the controversy
we have found remains. Before attempting to understand data using threshold 
resummation, our conclusion should be taken into account.

\vskip 1.0cm

This work was supported by the National Science Council of R.O.C. under
Grant Nos. NSC88-2112-M-006-013 and NSC88-2811-M007-026. We also thank National
Center for Theoretical Sciences in Taiwan for partial support.

\vskip 1.0cm

%\newpage

\end{document}